\documentclass[submission,copyright,creativecommons]{eptcs}
\providecommand{\event}{MBT 2012}

\usepackage{algorithmic}
\usepackage[]{algorithm}
\usepackage{graphicx}
\usepackage{amsthm}

\newtheorem{definition}{Definition}

\title{Constraint-Based Heuristic On-line Test Generation from Non-deterministic I/O EFSMs\thanks{This work was supported by the Estonian Science Foundation grant no. 7667 and ELIKO Competence Center.}}

\author{Danel Ahman\thanks{The first author was at the Tallinn University of Technology when the bulk of this work was carried out.}
\institute{Computer Laboratory\\ University of Cambridge\\ Cambridge, UK}
\email{danel.ahman@cl.cam.ac.uk}
\and
Marko K\"{a}\"{a}ramees
\institute{Department of Computer Science\\ Tallinn University of Technology\\
Tallinn, Estonia}
\email{marko.kaaramees@ttu.ee}
}
\def\titlerunning{Constraint-Based Heuristic On-line Test Generation from Non-deterministic I/O EFSMs}
\def\authorrunning{D. Ahman, \& M. K\"{a}\"{a}ramees}

\begin{document}

\maketitle

\begin{abstract}

We are investigating on-line model-based test generation from non-deterministic output-observable Input/Output Extended Finite State Machine (I/O EFSM) models of Systems Under Test (SUTs). We propose a novel constraint-based heuristic approach (Heuristic Reactive Planning Tester (\(\chi\)RPT)) for on-line conformance testing non-determinis\-tic SUTs. 
An indicative feature of \(\chi\)RPT is the capability of making reasonable decisions for achieving the test goals in the on-line testing process by using the results of off-line bounded static reachability analysis based on the SUT model and test goal specification.
 We present \(\chi\)RPT in detail and make performance comparison with other existing search strategies and approaches on examples with varying complexity. 

\end{abstract}

\section{Introduction}
\label{sec:introduction}

Model Based Testing (MBT) is one of various test automation approaches. We consider a version of MBT where the System Under Test (SUT) is represented by a formal model and treated as a "black-box" with an interface. MBT is commonly used to test conformance of the SUT to its model. A SUT may be modelled by a non-determistic model because of abstraction, distributed behaviour or freedom allowed in the specification. Testing conformance in that case requires on-line testing to react to the actual behaviour of the SUT.  

A widespread approach to modeling SUTs for test generation is using either Finite State Machines (FSMs) or  Extended Finite State Machines (EFSMs) \cite{lee1996,derderian2010a,vain2011}. Test generation that includes the input data generation from EFSM models has been handled with different methods, including evolutionary algrithms \cite{kalaji2008}, scenarios \cite{Veanes05OnlineTesting} and symbolic techniques \cite{vain2011}. The formal symbolic framework and notation of conformance for models involving data components is handled in \cite{frantzen2006,Veanes11AlternatingSimulation}. On-line methods for test generation from non-deterministic models have also been studied by various authors \cite{Nachmanson04optimalstrategies, Veanes06OnlineTesting,li2005, vain2011}. 

Although there is a variety of approaches for test generation from EFSMs, most of the methods are not applicable or tend to be inefficient when applied to non-deterministic and industrial-scale systems for on-line test generation for specified test goals (e.g., coverage criteria). In this paper we propose an Heuristic Reactive Planning Tester (\(\chi\)RPT) to improve the scalability and performance of Reactive Planning Tester \cite{vain2011,kaaramees2010} by an heuristic constraint-driven on-line test generation technique. The only approaches we are aware of that have comparable goals are presented in  \cite{derderian2010a,kalaji2009b}. The comparison is presented in Section \ref{sec:results}.

\begin{figure}
	\centering
		\includegraphics[width=9cm]{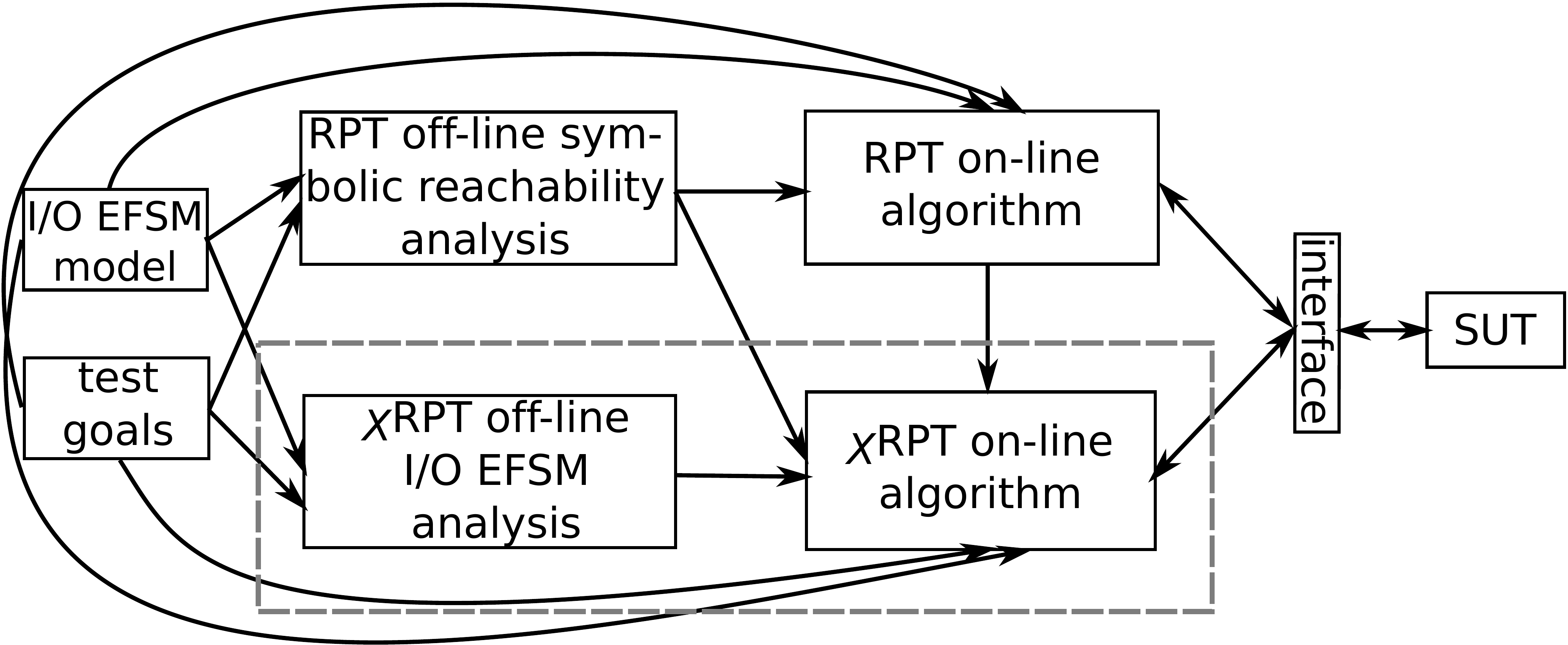}
	\caption{Workflow of I/O EFSM based on-line test generation and execution}
	\label{fig:workflow}
\end{figure}

The integral part of \(\chi\)RPT is an on-line decision-making algorithm responsible for computing the stimuli to the SUT based on various constraints emerging from the model of the SUT. This algorithm draws inspiration from the paradigm of Constraint-Based Local Search (CBLS) \cite{michel02aconstraint-based}, which has evolved into programming language Comet that we use for prototyping. As \(\chi\)RPT is based on on-line decision-making, it can also be used in reactive model-based planning in testing (cf. \cite{williams1997}). This is because \(\chi\)RPT computes only one move at a time and is able to react to the observed output of the SUT and the changes in test goals on-the-fly.
General workflow of I/O EFSM based on-line test generation and execution discussed in this paper is depicted in Fig. \ref{fig:workflow}.  The scope of \(\chi\)RPT is highlighted with a \emph{dashed rectangle}.

The rest of this paper is structured as follows. In Sect. \ref{sec:preliminaries} we outline the relevant background theory and preliminaries. Next, in Sect. \ref{sec:localsearch}, we describe \(\chi\)RPT in detail. The experimental results are described and analyzed in Sect. \ref{sec:results}. Finally, Sect. \ref{sec:discussion}  includes the discussion and further work.

\section{Preliminaries}
\label{sec:preliminaries}

In this section we introduce the background theory of \(\chi\)RPT. 
At first, in Sect. \ref{sec:efsm} and \ref{sec:testgoals}, we define the modeling formalism and general testing framework. Next, we give the necessary description of the Reactive Planning Tester (RPT) in Sect. \ref{sec:rpt}. The background theory is also illustrated with a simple three-variable counter example.

\subsection{Input/Output Extended Finite State Machines}
\label{sec:efsm}

In this paper we assume that the SUT is modelled as an output-observable deterministic or non-deterministic I/O EFSM over a first-order theory. For simplicity, the definitions given in this paper use formulas of first-order theory of linear integer arithmetic. It is also applicable to other theories where the Satisfiablity Modulo Theories (SMT) problem is decidable.

\begin{definition}
A constraint over variables \(X\) is a first-order formula of the chosen theory (e.g., over arithmetic expressions) where variables in  \(X\) occur as free variables. It is assumed, but not required to be quantifier-free for efficiency reasons.
\end{definition}

\begin{definition}
\label{def:ioefsm}
An I/O EFSM \(M\) is a tuple \((L, l_0, X, D, I, O, G, U, T)\) where \(L\) is a finite set of locations, \(l_0\) is an initial location, \(X = X_S \cup X_I \cup X_O \cup X_{Tr}\) is a disjoint set of finite sets of state, input, output and trap variables, \(D\) is a constraint over \(X\)  constraining the domain of variables, \(I\) is a finite set of input labels that may have an associated set of parameters \(x_1, ..., x_n\) \((x_k \in X_I)\), \(O\) is a finite set of output labels that may have an associated set of parameters \(x_1, ..., x_n\) \((x_k \in X_O)\), \(G\) is a finite set of guard conditions (constraints), \(U\) is a finite set of transition update functions and  \(T \subset L \times I \times O \times G \times U \times L\) is a finite set of transitions.
\end{definition}

The definition of I/O EFSMs is inspired by UML State-charts and allows intuitive modelling of interactions between the system and its environment. The main difference with the straightforward semantics of the formalisms of Symbolic Transition Systems (STSs) and Input-Output Labelled Transition Systems (IOL\-TSs) \cite{frantzen2006} is that I/O EFSMs allow transitions to have both input and output labels assigned to them simultaneously. It is possible to define the semantics based on STSs such that a transition of I/O-EFSM corresponds to two consecutive transitions of STSs. However, we keep the input and output  together because the presented method deals with interactions as unitary events. Guard conditions and all other constraints also implicitly include variable domain constraints \(D\) for all variables in \(X\).

Functions \(source(t)\), \(target(t)\), \(guard(t)\), \(update(t)\) and \(out(l) \subseteq T\) on I/O EFSMs serve as a short-hand reference to source and target location, guard condition and update function of a transition \(t\) and the set of outgoing transitions from location \(l\) respectively. 

\begin{definition}
A state \(s \in S\) of I/O EFSM is a pair \((l,\alpha)\) of a location \(l \in L\) and assignment \(\alpha\) of state variables in \(X_S\). 
\end{definition}

Therefore, input variables \(X_I\) and output variables \(X_O\) are not considered to be a part of an I/O EFSM state. The values of \(X_I\) and \(X_O\) are relevant only for the current transition. The input variables that model the parameters of input can only occur in the guards and in the right hand sides of  updates. Output variables that model the parameters of output can only occur in the left hand sides of updates.

A transition \((l, i, o, g, u, l')\) is  \emph{enabled} and can be taken when an input \(i\) is received and guard \(g\) evaluates to \(true\) on the current state and values of the input parameters. The receiving of input \(i\) is modelled on the logical level by a special input-variable \(iLabel\). We can say that a transition is enabled when a formula \(g  \wedge D \wedge iLabel=i\) evaluates to \(true\). 

\begin{definition}
An I/O EFSM is said to be non-deterministic, if there exists a state \(l\) for which two or more guard conditions of transitions in \(out(l)\) are non-disjoint and therefore satisfiable using the same input and state variables assignment. Transitions \(t \in out(l)\) satisfying this criteria are called rival transitions and are denoted \(Rival_t\). Transitions \(t' \in out(l)\) with \(guard(t')\) equal to or weaker than \(guard(t)\) (i.e., the guards are undistinguishable from each other for every assignment of input variables) are perfect rivals to \(t\) and denoted \(Rival^P_t\).
\end{definition}

It is further assumed that the SUT is modeled as an \emph{output observable} I/O EFSM. This means that even though in a given state \((l,\alpha)\) multiple rival transitions in \(out(l)\) may be taken in response to the input, the observed output of the SUT determines the actual move and the next state unambiguously. It is possible to relax the condition in expense of increasing the complexity of the on-line computation to find the best next move from a set of possible ones. We find this kind of limited non-determinism to be practical both for modelling and test generation point of view.

\begin{figure}
	\centering
		\includegraphics[width=9cm]{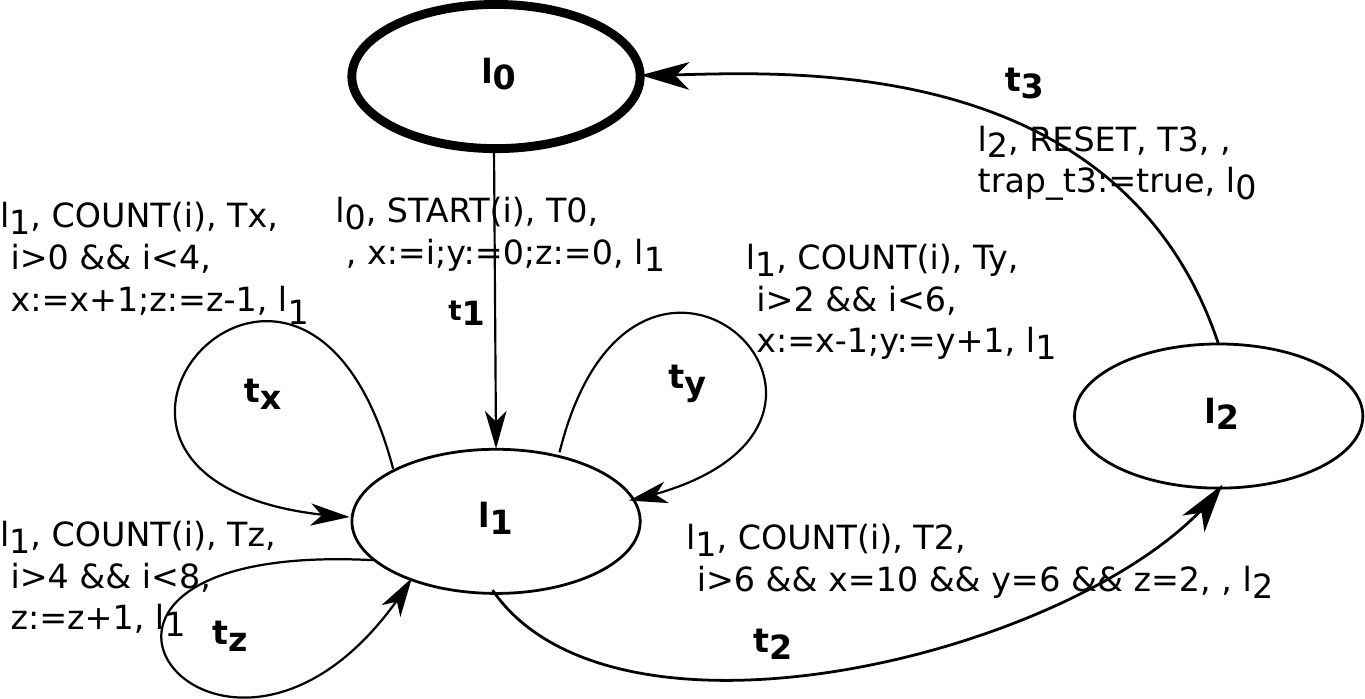}
	\caption{I/O EFSM model \(M_1\) of a simple three-variable counter}
	\label{fig:simpleloop11a}
\end{figure}

To further clarify the given notions, Fig. \ref{fig:simpleloop11a} describes an output observable non-deterministic I/O EFSM model of a simple thee-variable counter where all variables have domains [\(0,25\)]: \\ \(M_1=(L, l_0, X, D, I, O, G, U, T)\), where \(L=\{l_0,l_1,l_2\}\), \(X_S=\{x,y,z\}\), \(X_I= \{i\}\), \(X_O=\oslash\), \(X_{tr}=\{trap\_t_3\}\), \\ \(D=(0 \leq x \leq 25 \wedge 0 \leq y \leq 25 \wedge 0 \leq z \leq 25 \wedge 0 \leq i \leq 25 \wedge (trap\_t_3=true \vee trap\_t_3=false))\), \\ \(I=\{START,COUNT,RESET\}\),  \(O=\{T0,Tx,Ty,Tz,T2,T3\}\) \\  The model \(M_1\) is non-deterministic as it has the following sets of rival transitions: \(t_x\) with \(t_y\), \(t_y\) with \((t_x,t_z)\), \(t_z\) with \((t_y,t_2)\) and \(t_2\) with \(t_z\).

\subsection{Modeling Test Goals}
\label{sec:testgoals}

A \emph{test goal} is a property of the SUT that is intended to be tested. The test goals are modeled as sets of \emph{traps} attached to specific transitions. We classify traps as \emph{uncovered, covered} or \emph{discarded}.

\begin{definition}
\label{def:trap}
A trap is a pair \((t_i, P_{tr})\) where \(t_i\) is a transition and \(P_{tr}\) is a constraint on \( X_S \cup X_I \). The trap \((t_i, P_{tr})\) is covered when the transition \(t_i\) has been taken from the state and input where \(P_{tr}\) was satisfiable.
\end{definition}

In the following, a trap \((t_i, P_{tr})\) is denoted with a boolean \emph{trap variable} \(tr \in X_{Tr}\) (or sometimes with \(trap\_ti \in X_{Tr}\)). For a trap \((t_i, P_{tr})\), the value of the trap variable \(tr\) is \(false\) if the trap is either uncovered (initially all traps) or discarded and becomes \(true\) when the trap gets covered (cf. Def. \ref{def:trap}). 
An uncovered trap becomes discarded only when it is determined in the on-line algorithm that it can not be covered.

The constraint \(P_{tr}\) can also include trap variables \(tr'\) of other traps \((tr \neq tr')\) which, in turn, introduces a \emph{dependency relation} between the traps and enables one to talk about specific paths (i.e., sequences of trap-labeled transitions). Defining the test goals as sets of traps allows one to use different test strategies such as \emph{state, transition, path} and \emph{constrained path coverage} \cite{tahat2001}. Furthermore, we say that a test goal is \emph{fully satisfied} when all its traps have been covered.

\subsection{Reactive Planning Tester}
\label{sec:rpt}

The Reactive Planning Tester (RPT) is an on-line tester for black-box conformance testing of SUTs that are modeled by non-deterministic output-observable I/O EFSMs. As the RPT has been thoroughly described in \cite{vain2011}, we outline only the aspects relevant to \(\chi\)RPT.

Due to possible non-determinism, it is not possible to compute inputs for the SUT in advance, i.e., off-line. Therefore the workflow of the RPT is further divided into on-line and off-line procedures for efficient input generation and computationally hard symbolic analysis. The RPT off-line process performs symbolic reachability analysis and generates a system of reachability constraints describing the feasible paths (i.e., sequences of transitions) needed to be taken to cover the defined traps.

For every trap \(tr\) the RPT generates the following: 1) A \emph{weakest constraint} \(C_{l,tr}^*\) on state variables \(X_{S}\) and \emph{path length} \(\mathcal{L}_{l,tr}^*\) for every location \(l\). \(
C_{l,tr}^*\) represents a symbolic state 
for which there is a path with length (i.e., the number of transitions in a path) less or equal to \(\mathcal{L}^*_{l,tr}\) that covers the trap \(tr\).  2) A \emph{guarding constraint} \(C_{t,tr}^g\) on state and input variables \(X_S \cup X_I\) for every transition \(t\), which represents a symbolic state and input for which the transition \(t\) is the initial transition of a shortest path to the trap \(tr\). Assignment of \(false\) to any of the constraints represents a failure to generate a feasible constraint.

The reachability constraints are generated by a recursive procedure backwards from the trap until one of the following termination conditions is met - (i) a fixpoint is reached, (ii) the constraints have been generated for the initial location or (iii) a predefined bounded depth limit is reached. The case (iii) is common to location-, transition- or path-wise large systems where it is computationally infeasible to generate the constraints until termination condition (i) or (ii) is satisfied. 

During the on-line procedure, the goal of the RPT is to guide the SUT by using the reachability constraints generated off-line to satisfy the test goals. 
In general, the RPT on-line procedure works as follows. First, an uncovered trap with shortest path length \(\mathcal{L}_{l,tr}^*\)   and satisfiable constraint \(C_{l,tr}^*\) is selected in the current state \((l,\alpha)\). Then the assignment of input variables is computed by solving  \(C{}_{t,tr}^g\) (adding the negations of the guards of rival transitions where necessary) for transition \(t\) in \(out(l)\). Finally, the tester feeds the inputs to the SUT and observes its output to test conformance, simulates the transition and repeats the steps in the target location.

The RPT on-line algorithm is only applicable when the constraints have been generated for location \(l\) the SUT is in and \(C_{l,tr}^*\) is satisfiable in the given state \((l,\alpha)\). If either of these conditions is unsatisfied, the generated reachability constraints can not be directly used in this state. To overcome this problem, the RPT currently incorporates simple but inefficient random and anti-ant search strategies \cite{li2005} to make randomized intermediate moves  when the constraint based input generation is not possible. 

For illustrating the RPT, the reachability constraints generated for the model \(M_1\) in Fig. \ref{fig:simpleloop11a} and trap \((t_3, true)\) (denoted by $trap\_t3$, abbreviated as \(tr_3\)) using the bounded depth 2 are the following:  \((C^*_{l_0,tr_3}, \mathcal{L}^*_{l_0,tr_3}) \gets (false,-)\) , \((C^*_{l_1,tr_3}, \mathcal{L}^*_{l_1,tr_3}) \gets (x = 10 \wedge y = 6 \wedge z = 2,  2)\), \((C^*_{l_2,tr_3}, \mathcal{L}^*_{l_2,tr_3}) \gets (true,1)\),\\ \(C^g_{t_1,tr_3} \gets false\), \(C^g_{t_x,tr_3} \gets true\), \(C^g_{t_y,tr_3} \gets (x = 11 \wedge y = 5 \wedge z = 2)\), \(C^g_{t_z,tr_3} \gets (x = 10 \wedge y = 6 \wedge z = 1)\), \(C^g_{t_2,tr_3} \gets (x = 10 \wedge y = 6 \wedge z = 2)\) and \(C^g_{t_3,tr_3} \gets true\).

\section{Heuristic Reactive Planning Tester (\(\chi\)RPT)}
\label{sec:localsearch}

In this section we describe the Heuristic Reactive Planning Tester (\(\chi\)RPT) for I/O EFSM based on-line test generation. The aim of \(\chi\)RPT is to improve the scalability and efficiency of on-line test generation. \(\chi\)RPT is designed to be used when the RPT on-line algorithm is not directly applicable and is, similarly to the RPT, divided into off-line and on-line procedures described in  Sect. \ref{sec:offlinealg} and \ref{sec:onlinealg}. 

\subsection{Off-line Analysis of I/O EFSMs in \(\chi\)RPT}
\label{sec:offlinealg}

In this section we describe the off-line analysis of I/O EFSMs. The goal of this analysis is to provide extra input for the proposed on-line decision-making algorithm and to avoid unnecessary repetition of on-line computations. The analysis is made after the off-line symbolic reachability analysis in the RPT and prior to the on-line test generation and execution. This analysis is based on the I/O EFSM model \(M\) and the output generated by the RPT off-line process. As a result,  distance matrix \(Dist\), search neighbourhood \(L^C_{tr,l}\) and sets \(Tr_{+}\) , \(Tr_{-}\) of traps are generated as follows. 

The \emph{all-pairs shortest distance matrix} \(Dist\) is computed from the underlying directed control graph of the I/O EFSM with \(Dist_{l,l} = 0\) for reflexive transitions and \(Dist_{l_1,l_2} = \infty\) when there is no path from \(l_1\) to \(l_2\) for all \(l,l_1,l_2 \in L\). 
The use of graph based distances between locations is introduced to favour closer traps in the on-line decision making algorithm introduced in the next section.
 
The \emph{search neighbourhood} \(L^C\) is a set indexed both by traps \(tr \in X_{Tr}\) and locations \(l \in L\) of sets of closest (control graph based) locations (not necessarily direct neighbours) to \(l\) that have reachability constraints generated for them for a given trap \(tr\). If the location \(l \in L\) has reachability constraints generated for itself for trap \(tr\), \(L^C_{tr,l}\) includes both \(l\) and the next closest set of locations that have reachability constraints assigned to them for trap $tr$. In addition, the locations whose reachability constraints are equivalent to or weaker than domain constraints are removed from the search neighbourhood as they provide no information to our on-line algorithm. 
 
\(Tr_{+} \subset X_{Tr}\) and \(Tr_{-} \subset X_{Tr}\) are the respective \emph{sets of uncovered traps} that can be covered and that can not be covered from the current state. A function \(update\_neighbourhood(Tr_{+},Tr_{-})\) is used to update the two sets by removing already covered traps from \(Tr_{+}\) and moving new coverable traps from \(Tr_{-}\) to \(Tr_{+}\). In this paper it is assumed that \(M\) has a connected underlying control graph. As a result, the partitioning of traps between \(Tr_{+}\) and \(Tr_{-}\) in function \(update\_neighbourhood(Tr_{+},Tr_{-})\) is based only on the dependency ordering of the traps, i.e., which trap variables are used in the constraints of other traps. After the off-line analysis has been completed, \(Tr_{+}\, \cup\, Tr_{-} = Tr\) holds. 

In case the underlying control graph of \(M\) not being connected, one could also add a strongly connected component (SCC) analysis to the algorithm. This could then be used to give higher priority to traps in the current SCC to cover them before moving to the next SCC. Moreover, one could also add other selection and partitioning criteria but this is left as a further research.

\subsection{On-line Decision-Making Algorithm of \(\chi\)RPT}
\label{sec:onlinealg}

In this section we describe the on-line algorithm of \(\chi\)RPT responsible for decision-making during on-line test generation in situations when the RPT on-line algorithm is not immediately applicable (cf. discussion in Sect. \ref{sec:rpt}). The \(\chi\)RPT on-line algorithm consists of a top-level algorithm in Fig. \ref{fig:onlinealg} and four subroutines in Figs. \ref{fig:onlinealg2}, \ref{fig:onlinealg3}, \ref{fig:onlinealg4}, \ref{fig:onlinealg5}. They all make use of the I/O EFSM model \(M\), the reachability constraints generated by the RPT off-line algorithm and the output of \(\chi\)RPT off-line algorithm discussed in Sect. \ref{sec:offlinealg}.

\begin{figure}
	\begin{algorithmic}[1]
		\STATE \textbf{\textsc{On\_Line\_Algorithm}}(\(M, l, \alpha, X, D, L^C, L^T, Dist, Tabu, C^*, Tr_{+}, Tr_{-}\)):
		\vspace{0.1cm}
		\STATE \textbf{while} \(Tr_{+} \neq \emptyset\) \textbf{do}
		\STATE \hspace{0.3cm} $(Tr_{+},Tr_{-}) \gets $ \textsc{On\_Line\_RPT}(\(l, \alpha, C^*, Tr_{+}, Tr_{-}\))
		\STATE \hspace{0.3cm} $(MinHeap, Tr_{+},Tr_{-}, L^T) \gets $ 
		\STATE \hfill \textsc{Generate\_Solution\_Candidates}(\(l, \alpha, X, D, L^C, L^T, Dist, Tabu, C^*, Tr_{+}, Tr_{-}\))
		\STATE \hspace{0.3cm} $best\_move \gets $ \textsc{Choose\_Most\_Promising}(\(l, \alpha, X, D, L^C_{tr,l}, Dist, Tabu, C^*, Tr_{+}, Tr_{-}\))
		\STATE \hspace{0.3cm} $new\_state \gets $  \textsc{Interact\_With\_SUT}(\(M, l, \alpha, best\_move\))
		\STATE \hspace{0.3cm} \textbf{if} $new\_state = ()$ \textbf{then}
		\STATE \hspace{0.6cm} \textbf{return TEST\_FAILED}
		\STATE \hspace{0.3cm} \textbf{else} 
		\STATE \hspace{0.6cm} $(Tabu, l, \alpha) \gets new\_state$
		\STATE \textbf{return TEST\_FINISHED}
	\end{algorithmic}
	\caption{On-line decision-making algorithm of $\chi$RPT consisting of four subroutines}
	\label{fig:onlinealg}
\end{figure}

\begin{figure}
	\begin{algorithmic}[1]
		\STATE \textbf{\textsc{On\_Line\_RPT}}(\(l, \alpha, C^*, Tr_{+}, Tr_{-}\)):
		\vspace{0.1cm}
		\STATE \hspace{0.0cm} \textbf{while} \(\pi_1 (\textsc{SAT\_Model}(X_I, \alpha, C^*_{l,tr}))\) \textbf{for any} \(tr \in Tr_+\) \textbf{then}
		\STATE \hspace{0.3cm} \((l, \alpha) \gets \textsc{RPT\_On\_Line\_Algorithm}(tr,l, \alpha)\)
		\STATE \hspace{0.3cm} \textsc{Set\_Covered}$(tr)$ ; \textsc{Update\_Neighbourhood}$(Tr_{+},Tr_{-})$
		\STATE \hspace{0.0cm} \textbf{return} $(Tr_{+},Tr_{-})$
	\end{algorithmic}
	\caption{Subroutine \#1 that uses the RPT on-line algorithm to cover a given trap as soon as reachability constraints are satisfied} 
	\label{fig:onlinealg2}
\end{figure}

\begin{figure}
	\begin{algorithmic}[1]
		\STATE \textbf{\textsc{Generate\_Solution\_Candidates}}(\(l, \alpha, X, D, L^C, L^T, Dist, Tabu, C^*, Tr_{+}, Tr_{-}\)):
		\vspace{0.1cm}
		\STATE \hspace{0.0cm} \textbf{for all} \(run \in \{0,1\}\) \textbf{do}
		\STATE \hspace{0.3cm} \textbf{for all} \(tr \in Tr_+\) \textbf{do}
		\STATE \hspace{0.6cm} $MinHeap' \gets \emptyset$
		\STATE \hspace{0.6cm} \textbf{for all} \(t \in out(l)\) \textbf{do}
		\STATE \hspace{0.9cm} \(formula \gets guard(t) \bigwedge_{t' \in Rival_t} \lnot guard(t') \wedge D(X)[update(t)/X]\)
		\STATE \hspace{0.9cm} \(formula \gets formula \wedge \lnot Tabu_{tr,l}\)
		\STATE \hspace{0.9cm} \((b , \alpha_i) \gets \textsc{Sat\_Model}(t,X_I, \alpha, formula)\)
		\STATE \hspace{0.9cm} \textbf{if} \(\lnot b\) \textbf{then} \textbf{continue}
		\vspace{0.1 cm}
		\STATE \hspace{0.9cm} \textbf{for all} \(l_C \in L^C_{tr,l}\) \textbf{do}
		\STATE \hspace{1.2cm} \textbf{if} \((run = 0) \lor (run = 1 \wedge \lnot (target(t) \in L^T_{tr} \wedge source(t) \in L^T_{tr} ))\) \textbf{then}
		\STATE \hspace{1.5cm} \(dist \gets 1 + Dist_{target(t),l_C}\)
		\STATE \hspace{1.5cm} \(viol \gets \nu(C^*_{l_C,tr}[update(t)/X][\alpha(x_s)/x_s][\alpha_i(x_i)/x_i])\)
		\STATE \hspace{1.5cm} \(f \gets dist^2 + viol^2\)
		\STATE \hspace{1.5cm} \(MinHeap' \gets  MinHeap' \cup (t, \alpha_i, l_C, tr, f)\)
		\STATE \hspace{0.6cm} \textbf{if} \(MinHeap' = \emptyset\) \textbf{then}  
		\STATE \hspace{0.9cm} \textbf{if} run = 0 \textbf{then}
		\STATE \hspace{1.2cm} \(Tabu_{tr,l} \gets \emptyset\) ; \(L^T_{tr} \gets L^T_{tr} \cup l\)
		\STATE \hspace{0.9cm} \textbf{else}
		\STATE \hspace{1.2cm} \textsc{Set\_Discarded}\((tr)\) ; \textsc{Update\_Neighbourhood}\((Tr_{+},Tr_{-})\)
		\STATE \hspace{0.6cm} \textbf{else} 
		\STATE \hspace{0.9cm} $MinHeap \gets MinHeap \cup MinHeap'$
		\STATE \hspace{0.3cm} \textbf{if} $MinHeap \neq \emptyset$ \textbf{then}
		\STATE \hspace{0.6cm} \textbf{break}
		\STATE \hspace{0.0cm} \textbf{return} $(MinHeap, Tr_{+},Tr_{-}, L^T)$
	\end{algorithmic}
	\caption{Subroutine \#2 that generates solution candidates, excludes excessive ones and orders the remaining by fitness function values} 
	\label{fig:onlinealg3}
\end{figure}

\begin{figure}
	\begin{algorithmic}[1]
		\STATE \textbf{\textsc{Choose\_Most\_Promising}}(\(l, \alpha, X, D, L^C_{tr,l}, Dist, Tabu, C^*, Tr_{+}, Tr_{-}\)):
		\STATE \hspace{0.0cm} \(best\_f = \infty\) ; \(best\_move = \emptyset\)
		\STATE \hspace{0.0cm} \textbf{for up to} \(N\) \textbf{tuples} \((t, \alpha_i', l_C, tr, f) \in MinHeap\) \textbf{do}
		\STATE \hspace{0.3cm} \(formula \gets guard(t) \wedge D(X)[update(t)/X] \wedge \lnot Tabu_{tr,l} \)
		\STATE \hspace{0.3cm} \(formula \gets formula \bigwedge_{t' \in Rival_t} \lnot guard(t')\)
		\STATE \hspace{0.3cm} \((b , \alpha_i) \gets \textsc{Optimize\_Model}(t,X_I, \alpha, formula , \nu(C^*_{l_C,tr}[update(t)/X]))\)
		\vspace{0.1 cm}
		\STATE \hspace{0.3cm} \(dist \gets 1 + Dist_{target(t),l_C}\)
		\STATE \hspace{0.3cm} \(viol \gets \nu(C^*_{l_C,tr}[update(t)/X][\alpha(x_s)/x_s][\alpha_i(x_i)/x_i])\)
		\STATE \hspace{0.3cm} \(f \gets dist^2 + viol^2\)
		\vspace{0.1 cm}
		\STATE \hspace{0.3cm} \textbf{if} \(f < best\_f\) \textbf{then} 
		\STATE \hspace{0.6cm} \(best\_f \gets f\)
		\STATE \hspace{0.6cm} \(best\_move \gets (\alpha_i, t, l_C, tr, f)\)
		\vspace{0.1 cm}
		\STATE \hspace{0.0cm} \textbf{return} $best\_move$
	\end{algorithmic}
	\caption{Subroutine \#3 that selects the most promising solution candidate as a best possible move from a given state} 
	\label{fig:onlinealg4}
\end{figure}

\begin{figure}
	\begin{algorithmic}[1]
		\STATE \textbf{\textsc{Interact\_With\_SUT}}(\(M, l, \alpha, best\_move\)):
		\vspace{0.1 cm}
		\STATE \hspace{0.0cm} $(\alpha_i, t, l_C, tr, f) \gets best\_move$
		\STATE \hspace{0.0cm} $iLabel \gets \textsc{Get\_ILabel}(\alpha_i)$
		\STATE \hspace{0.0cm} \(actual\_move \gets \textsc{Feed\_To\_SUT}(\textsc{Create\_Message}(iLabel, \alpha_i)))\) 
		\STATE \hspace{0.0cm} \textbf{if} \(\textsc{Sut\_Conforms}(M,actual\_move)\) \textbf{then}
		\STATE \hspace{0.3cm} \(Tabu_{tr,l} \gets Tabu_{tr,l} \vee \textsc{Make\_Tabu\_Element}(actual\_move, best\_move)\)
		\STATE \hspace{0.3cm} $(l, \alpha) \gets \textsc{Get\_State}(\textsc{Simulate\_Move}(actual\_move))$
		\STATE \hspace{0.3cm} \textbf{return} $(Tabu, l, \alpha)$
		\STATE \hspace{0.0cm} \textbf{else} 
		\STATE \hspace{0.3cm} \textbf{return} $()$
	\end{algorithmic}
	\caption{Subroutine \#4 that creates a message based on the best move, feeds this message to the SUT and observes its output} 
	\label{fig:onlinealg5}
\end{figure}

\begin{figure}
	\begin{center}
		\begin{tabular}{l c l}
			\(A, B - logical \ formulas \) & \ \ \ \ \ \ \ \ \  & \(\nu(a >= b) = abs(min(0, \nu(a) - \nu(b)))\) \\
			\(a, b - arithmetic\ expressions\) & & \(\nu(a > b) = abs(min(0, -1 + \nu(a) - \nu(b)))\) \\
			\(\nu(a = b) = abs(\nu(a) - \nu(b))\) & & \(\nu(a < b) = abs(max(0, 1 + \nu(a) - \nu(b)))\) \\
			\(\nu(a \neq b) = \nu(a < b \vee a > b)\) & & \(\nu(a <= b) = abs(max(0, \nu(a) - \nu(b)))\) \\
			\(\nu(A \wedge B) = \nu(A) + \nu(B)\) & & \(\nu(A \vee B) = min(\nu(A), \nu(B))\)
  		\end{tabular}
	\end{center}
\caption{Minimal set of computation rules for the violations degree function \(\nu\) 
}
\label{fig:violationsdegreefunction}
\end{figure}

In general, the on-line algorithm of $\chi$RPT works by making computationally inexpensive operations first and then iteratively excludes \emph{solution candidates} (i.e., possible moves) as the computations become more costly until the most promising solution candidate has been selected. \emph{Solution candidates} are tuples \((t, \alpha_i, l_C, tr, f)\) consisting of a transition $t$, input variables assignment $\alpha_i$, search neighbourhood location $l_C$, trap $tr$ and fitness function value $f$ which is used to measure and compare the quality of solution candidates. In this paper, the \emph{fitness function} consists of the sum of squares of the control-graph based distance to the search neighbourhood location \(l_C\) and the violations degree (cf. Def. \ref{def:violationsdegree}) of the reachability constraint \(C_{l_C,tr}^*\). The $\chi$RPT on-line algorithm also makes use of the RPT on-line algorithm when the SUT has been guided to a state where reachability constraints for at least one trap are satisfied making the RPT on-line algorithm  applicable. The four subroutines that the $\chi$RPT on-line algorithm (Fig. \ref{fig:onlinealg}) consists of are described in the following paragraphs.

\begin{definition}
\label{def:violationsdegree}
The violations degree of a constraint \(C\) is the value of the function \(\nu(C)\) that is inspired by fitness function computation in \cite{tracey98a}. The minimal set of computation rules for \(\nu(C)\) is given in Fig. \ref{fig:violationsdegreefunction}.
The negation of logical formulas is pushed inside and eliminated (if possible) by De Morgan's laws and arithmetic equivalences (i.e., \(\nu(\lnot(a > b)) \equiv \nu(a <= b)\)).
\end{definition}

\textbf{Subroutine \#1} The $\chi$RPT on-line algorithm uses the RPT on-line algorithm through the subroutine $\textsc{On\_Line\_RPT}(...)$ outlined in Fig. \ref{fig:onlinealg2} as soon as the SUT has been guided to a state \((l,\alpha)\) where reachability constraints \(C_{l,tr}^*\) for some trap \(tr\) are satisfied. Then, the RPT on-line algorithm is called using the procedure \(\textsc{RPT\_On\_Line\_Algorithm}(l, \alpha,\) \(tr)\) to guide the SUT to a state \((l',\alpha')\) covering \(tr\). Next, if \(C_{l',tr'}^*\) is satisfied for any \(tr'\) in the state \((l',\alpha')\) such that \(tr \neq tr'\), the routine is repeated. 

\textbf{Subroutine \#2} The subroutine $\textsc{Generate\_Solutions\_Candidates}(...)$ in Fig. \ref{fig:onlinealg3} is used to generate a limited amount of solution candidates by excluding excessive ones. The core idea of this routine is to collect solution candidates ordered by the calculated fitness using the min-heap \(MinHeap\). Excessive solution candidates are excluded by strengthening the guards of transitions \(t\) in \(out(l)\) for the \emph{satisfiability test} procedure \(\textsc{SAT\_Model}(t,X_I, \alpha, formula)\) with the negations of guards of rival transitions and \emph{tabu list} \(Tabu_{tr,l}\) element (ll. 6-7). Given the assignment \(\alpha\) of state variables, \(\textsc{SAT\_Model}(t,X_I, \alpha, formula)\) returns a pair \((b,\alpha_I)\) of a boolean value indicating whether the logic formula was satisfiable and a model for variables in \(X_I\). The algorithm uses a \emph{tabu list} to avoid converging into local optimums and therefore avoid guiding the SUT to infinite loops by keeping the partial history of previous moves.
For every trap \(tr\) and location \(l\), the tabu list element \(Tabu_{tr,l}\) consists of a disjunction of conjunctions of I/O EFSM transition together with input and state variables assignments. These elements explicitly record the moves made in the on-line algorithm. Tabu list becomes \emph{full} if none of the guards of \(t \in out(l)\), strengthened with the negations of tabu list elements, are satisfiable in the given state. The general  principles of tabu lists and related search strategies can be found from papers by Fred Glover et al. (e.g., \cite{glover1993}).

If no solution candidates for trap \(tr\) are collected to \(MinHeap\) on the first run, then the guards are weakened by emptying the tabu list \(Tabu_{tr,l}\). At the same time, the location \(l\) is added to \(L^T_{tr}\) (the set of locations \(l\) for each trap \(tr\) where tabu list \(Tabu_{tr,l}\) has been emptied). In addition, in the second run, a new condition is added (l. 11) to avoid infinite looping. This condition states that if the tabu lists in the current location \(l\) and target location of transition \(t \in out(l)\) have been emptied before (i.e., \(target(t) \in L^T_{tr} \wedge source(t) \in L^T_{tr}\)) then the move is not permitted. If no solution candidates are found for trap \(tr\) after the second run due to the new condition, then \(tr\) is marked as discarded. Unfortunately this condition does not guarantee the complete unreachability of \(tr\) but is merely an over-approximation which turns out to be strong enough for \(\chi\)RPT. Therefore the notion \textit{discarded} is used instead of \textit{unreachable}.

\textbf{Subroutine \#3} The subroutine $\textsc{Choose\_Most\_Promising}(...)$ in Fig. \ref{fig:onlinealg4} is used to choose the most promising solution candidate (i.e., the best possible move in a given state). This subroutine compares only up to \(N \in \{1,2,...,\) \(size(MinHeap)\}\) solution candidates  from all the solution candidates collected in \(MinHeap\). We allow to vary \(N\) to allow different configurations to be used, e.g., for different time requirements.
However, the selection cost in this round is higher than before because of the use of \emph{constraint solving} in procedure \(\textsc{Optimize\_}\) \(\textsc{Model}(t,X_I, \alpha, C, f)\) (as opposed to SAT test used in Subroutine \#2), which also optimizes the assignment \(\alpha_i\) of input values to minimize  the fitness function \(f\). 

\textbf{Subroutine \#4} The subroutine $\textsc{Interact\_With\_Sut}(...)$ in Fig. \ref{fig:onlinealg5} is used for interaction with the SUT using the most promising solution candidate $best\_move$ found in Subroutine \#3. The message that will be fed to the SUT consists of an input label together with possibly empty list of parameters (cf. Def. \ref{def:ioefsm}). The input label is obtained from the valuation of the variable $iLabel$. The input label also determines the input variables whose valuation will be sent as input parameters. After feeding this input message to the SUT, the subroutine observes the actual move made. If the SUT conforms to the I/O EFSM model, a tabu list element \(Tabu_{tr,l}\) is updated with the result of \(\textsc{Make\_Tabu\_Element}(actual\_move, best\_move)\) that is a constraint recording the actual or the best move (cf.  comments below). Finally, the algorithm returns the updated tabu list and the new state corresponding to making \(actual\_move\). 

It has to be noted that if one would only consider actual moves made by the SUT for tabu list element construction, then non-determinism of perfect rivals might force the algorithm to loop infinitely. The \(actual\_move\) and \(best\_move\) need not be the same and the tabu list would not reflect the history correctly. Therefore, in our algorithm, we enforce that \(best\_move\) is used for tabu list element construction if \(actual\_move\) is already present in the tabu list. Instead of this, a more sophisticated bounded fairness criteria could be introduced but it has been omitted from this paper due to space restrictions.
 
Using \(\chi\)RPT has two possible outcomes. First, testing can be declared \emph{failed} if the observed behavior of the SUT does not conform to the given I/O EFSM model. Alternatively, testing is declared \emph{finished} when \(Tr_{+}\) becomes empty and no uncovered traps can be added to it. At this point, not all of the test goals might be satisfied because some of the traps might still be discarded or uncovered. One should then add these traps back to \(Tr_{-}\), reset the SUT and run the on-line algorithm again from the initial state.

The decision-making time of this algorithm is dictated by \(\textsc{SAT\_Model}\) and \(\textsc{Optimize\_Model}\) as these are the two most costly operations (in the worst case double exponential to the size of constraints). The size of the constraints depends non-trivially on the structure of the I/O EFSM model, RPT planning depth limit and simplifications involved. In \(\chi\)RPT, the number of calls to these operations in each state \((l,\alpha)\) is linear to the number of traps, locations in \(L^C_{tr,l}\) and transitions in \(out(l)\). Therefore, the performance could be risen by considering different heuristic (sub)methods for the most promising solution candidate selection to reduce the number of calls made to these operations. The analysis of such heuristic algorithms (e.g., simulated annealing or differential evolution) is omitted from this paper.  In the worst case, \(\chi\)RPT falls back to an anti-ant-like strategy and has similar performance. On the other hand, experimental results in the next section give evidence that, on average, \(\chi\)RPT is superior when compared to anti-ant by offering significantly better performance and measures to ensure termination.

We continue with the three-variable counter example \(M_1\) (we abbreviate $trap\_t3$ as $tr_3$) to illustrate \(\chi\)RPT further. First, we look at \(\chi\)RPT in the initial state \((l_0, \{x \gets 0, y \gets 0, z \gets 0\})\) where there is only one solution candidate as the only transition in \(out(l_0)\) is \(t_1\) and the only location in \(L^C_{tr_3,l_0}\) is \(l_1\). In this situation, \(\textsc{SAT\_Model}\) returns us a tuple \((true, i = 0)\) and thus the value of the fitness function is \(f \gets 1^2 + 18^2\). On the other hand, \(\textsc{Optimize\_Model}\) minimizes the value of \(f\) and returns \((true, i = 10)\) and thus \(f \gets 1^2 + 8^2\). 

Secondly, we look at \(\chi\)RPT in the next state \((l_1, \{x \gets 10, y \gets 0, z \gets 0\})\). This time we have four transitions \(t_x,t_y,t_z,t_2\) in \(out(l_1)\) and one location \(l_1\) in \(L^C_{tr_3,l_1}\). \(\textsc{SAT\_Model}\) first eliminates solution candidates for transitions \(t_x\) (as \(z\) would violate domain constraints) and \(t_2\) (as the guard is not satisfied). On the other hand, both \(\textsc{SAT\_Model}\) and \(\textsc{Optimize\_Model}\) return \((true, i = 4)\) and \((true, i = 6)\) for other solution candidates corresponding to \(t_y, t_z\). As a result, the fitness function values are  \(f_y \gets 1^2 + 8^2\) and \(f_z \gets 1^2 + 7^2\), and therefore, the solution candidate corresponding to \(t_z\) is selected as the most promising. This process continues until the state \((l_1, \{x \gets 10, y \gets 6, z \gets 2\})\) is reached where the reachability constraint \(C^*_{l_1,tr_3}\) is satisfied and the RPT on-line algorithm can be applied to cover trap \(tr_3\).

\section{Experimental Results}
\label{sec:results}

In this section we compare different strategies for generating test sequences for specific test goals. In particular, we compare the performance of \(\chi\)RPT with other search strategies such as the RPT off-line algorithm \cite{kaaramees2010,vain2011} and the  randomized version of the anti-ant strategy \cite{li2005}. 
As  $\chi$RPT can be viewed as heuristic explicit-state forward reachability analysis
rather than the symbolic analyzis done in the RPT, we have also chosen an explicit-state model checking tool UPPAAL \cite{larsen1997} with modelling language close to EFSM for comparison. It gives a comparison between the guided (\(\chi\)RPT) and random (UPPAAL) explicit-state forward analysis. Although our method is intended for non-deterministic models, the comparison is easier to make on deterministic models.
The following experiments were conducted on a 64-bit personal computer with 2.4GHz Intel Core 2 Duo CPU and 8GB of DDR3 RAM using prototypes written in Comet. 

\vspace{-0.6em}
\subsection{Single Trap Test Goals}
\label{sec:onetrap}

\begin{figure}
	\centering
		\includegraphics[width=13cm]{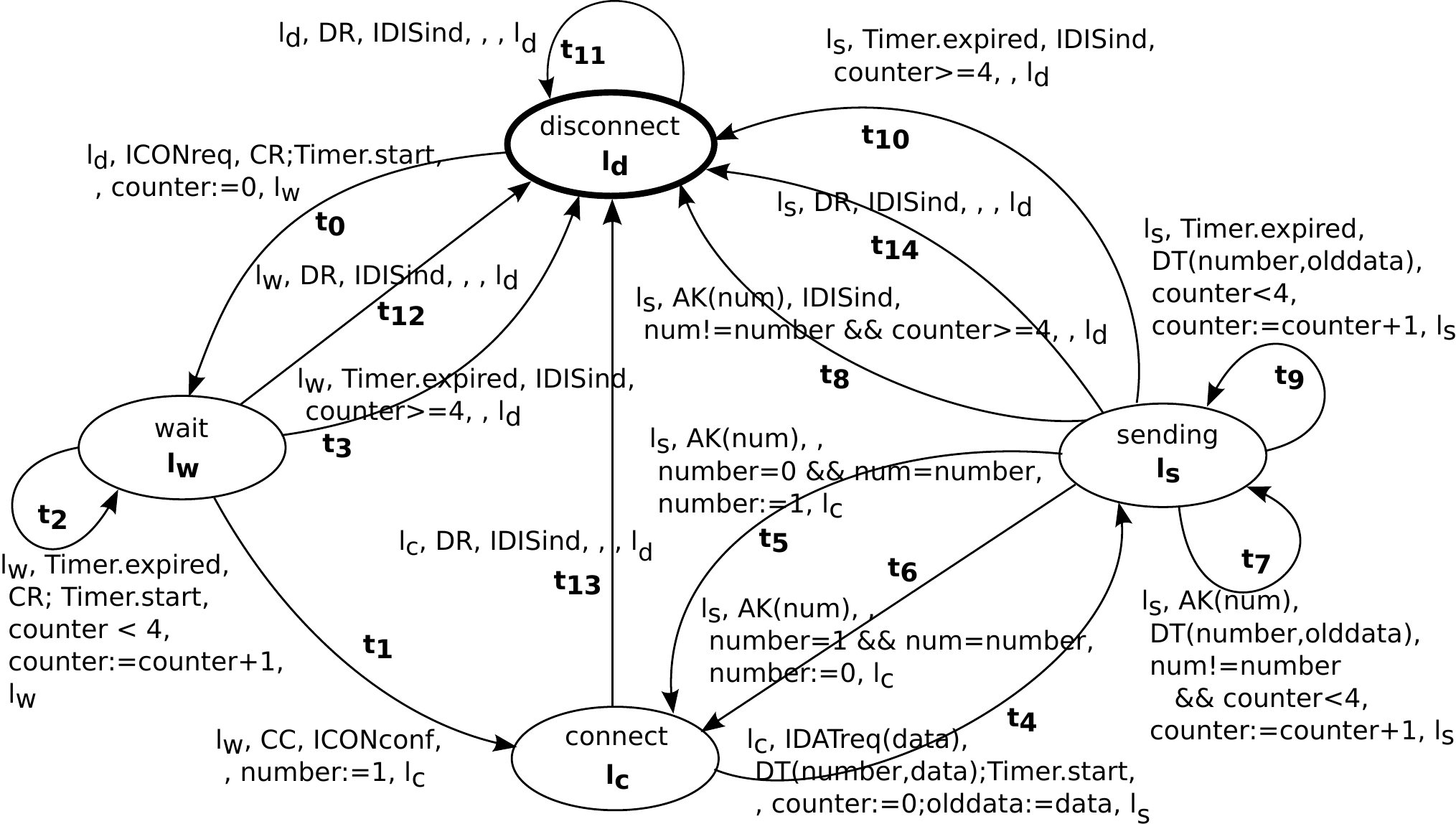}
	\caption{I/O EFSM model \(M_2\) of Inres Initiator}
	\label{fig:InresProtocol}
\end{figure}

In this section we analyse the three-variable counter introduced in Fig. \ref{fig:simpleloop11a} and the Inres Initiator depicted in Fig. \ref{fig:InresProtocol}. We consider only test goals containing one trap to give evidence of the performance of \(\chi\)RPT. 

The Inres protocol is a well-known case study model in software testing and verification communities. The connection-oriented protocol  consists of an Initiator that sets up a connection and sends data and a Responder that receives the data and closes the connection. In this paper we consider only the Inres Initiator depicted in Fig. \ref{fig:InresProtocol} which mimics the formalization given in \cite{derderian2010a}.

Experimental results in Table \ref{tab:onetrapresults} describe the performance of \(\chi\)RPT on three different test goals. One of the goals is defined on the three-variable counter model \(M_1\) consisting of trap \(trap\_t_3 = (t_3, true)\). The other two are defined on the Inres Initiator model \(M_2\) - \(trap\_t_8 = (t_8, true)\) and \(trap\_t_5 = (t_5, true)\). The respective initial states of I/O EFSM models \(M_1\) and \(M_2\) are \((l_0, \{x \gets 0, y \gets 0, z \gets 0\})\) and \((l_d, \{number \gets 0, counter \gets 0\})\). For all three traps, the reachability constraints are generated to depth 2 by the bounded RPT off-line algorithm. 
The results in Table \ref{tab:onetrapresults} outline the optimal length of paths generated by the RPT off-line algorithm and UPPAAL together with algorithm run times. The results also describe the length of paths generated by $\chi$RPT (completing the path with the RPT on-line algorithm as soon as it is possible) and the paths generated by the randomized anti-ant search strategy.

\begin{table}
  \caption{Experimental results of test goals consisting of one trap showing the lengths of the generated paths and algorithm run times}
	\begin{center}
		\begin{tabular}{| l | | c | c | c | c |}
			\hline
			 & \(M_1\) \(trap\_t_3\) & \(M_2\) \(trap\_t_8\) & \(M_2\) \(trap\_t_5\)\\
			\hline
			\textbf{Complete off-line RPT} & & &\\
			Path length (Work time (s)) & 11 (14.98) & 8 (6.84) & 6 (6.49) \\
			\hline
			\textbf{Bounded off-line RPT} & & & \\
			Depth (Work time (s)) & 2 (2.09) & 2 (4.5) & 2 (4.5)\\
			\hline
			\textbf{\(\chi\)RPT}  & & & \\
			Total path length (Work time (s)) & 23 (0.20) & 8 (0.043) & 6 (0.037) \\
			\hline
			On-line path length ($\chi$RPT + RPT) & 21 + 2 & 6 + 2 & 4 + 2 \\
			\hline
			Time spent in each state (s) & 0.0082 & 0.0061 & 0.0074\\
			\hline
			\textbf{Randomized anti-ant} & & &  \\
			Work time (s) (min/avg/max) &  - & 0.064\, /\, 0.35\, /\, 0.88  & 0.017\, /\, 0.11\, /\, 0.21\\
			\hline
			Path length (min/avg/max) & -  & 17\, /\, 80\, /\, 193 & 6\, /\, 25\, /\, 48\\
			\hline
			\textbf{UPPAAL} & & & \\
			Path length (Work time (s)) & 11 (0.53) & 8 (0.50) & 6 (0.49)\\
			\hline
  		\end{tabular}
	\end{center}
  \label{tab:onetrapresults}
\end{table}

The experimental results first indicate that \(\chi\)RPT is able to satisfy all three test goals. Moreover, it also outperformed the randomized anti-ant strategy significantly in all three cases. The path generated by \(\chi\)RPT is optimal in case of \(trap\_t_8\) and \(trap\_t_5\), but differs from the optimal computed by the RPT off-line algorithm and UPPAAL in case of \(trap\_t_3\). The difference is not significant when compared to the failure of the anti ant search strategy and is caused by the combination of assignments in \(update(t_1)\) and the optimization of input variables assignment. In Sect. \ref{sec:billingsystem} we also show that the difference between optimal and generated paths stays insignificant in case of  larger industrial systems.

\(\chi\)RPT also performed efficiently time-wise. In all of the three test goals, the average time spent on decision making in each state is well below 10ms making \(\chi\)RPT a feasible candidate for industrial on-line testing frameworks. In addition, it is clearly visible that the combination of $\chi$RPT on-line algorithm together with bounded RPT off-line algorithm also time-wise outperformed the RPT (off-line + on-line).

\vspace{-0.6em}
\subsection{Multiple Trap Test Goals}
\label{sec:multipletraps}

In this section we analyse the paths generated by \(\chi\)RPT for test goals consisting of multiple traps defined on the Inres Initiator model \(M_2\) (Fig. \ref{fig:InresProtocol}). All of the traps considered here are dependent on preceding traps which means the traps can only be covered in the order they are defined. The reachability constraints are generated to depth 1 by the bounded RPT off-line algorithm and the initial state is taken from earlier.

Many of the test goals in Table \ref{tab:multtrapresults} are inspired and partially taken from \cite{derderian2010a,kalaji2008}. We consider both (i) traps that can be covered immediately one after another and (ii) traps that can not. In particular, traps of form (ii) force the SUT to be guided through a set of intermediate states before the next trap can be covered. Here, the definition of Inres Initiator given in  \cite{derderian2010a} is used. The definition in \cite{kalaji2008} differs from the former by \(t_0\) and \(t_4\) and causes some of the test goals from \cite{kalaji2008} initially consisting of traps of form (i) to be actually of the form (ii).

\begin{table}
  \caption{Generated paths (of the EFSM model transitions) for test goals consisting of multiple traps defined on the Inres Initiator model. Lists of transitions in parentheses illustrate how the SUT is guided through intermediate states to cover the next trap}
	\begin{center}
		\begin{tabular}{| l| l | l |}
			\hline
			No. & Test goal & Generated path\\
			\hline
			1 & \(t_0-t_1-t_4-t_7-t_6-t_4-t_5-t_4\) & \(t_0-t_1-t_4-t_7-t_6-t_4-t_5-t_4\)\\
			\hline
			2 & \(t_{11}-t_0-t_1-t_4-t_6-t_4-t_5-t_4\) & \(t_{11}-t_0-t_1-t_4-t_6-t_4-t_5-t_4\)\\
			\hline
			3 & \(t_0-t_2-t_1-t_4-t_7-t_6-t_4-t_5\) & \(t_0-t_2-t_1-t_4-t_7-t_6-t_4-t_5\)\\
			\hline
			4 & \(t_0-t_3-t_0-t_1-t_4-t_6-t_4-t_7\) & \(t_0-(t_2,t_2,t_2,t_2,t_3)-t_0-t_1-t_4-t_6-t_4-t_7\)\\
			\hline
			5 & \(t_0-t_2-t_1-t_4-t_7-t_7-t_7\) & \(t_0-t_2-t_1-t_4-t_7-t_7-t_7-(t_9,t_8)\)\\
			\hline
			6 & \(t_1-t_8-t_{13}-t_5-t_{14}-t_2-t_9-t_{11}\) & \((t_0,t_1)-(t_4,t_9,t_9,t_9,t_9,t_8)-(t_0,t_1,t_{13})-\) \\ 
			& & \((t_0,t_1,t_9,t_9,t_9,t_9,t_6,t_4,t_5)-
			(t_4,t_{14})-(t_0,t_2)-\)\\
			& & \((t_1,t_4,t_9)-(t_{14},t_{11})\)\\
			\hline
			7 & \(t_2-t_{10}-t_6-t_3-t_4-t_{11}-t_7-t_0\) & \((t_0,t_2)-(t_1,t_4,t_9,t_9,t_9,t_9,t_{10})-(t_0,t_1,t_4,t_6)-\) \\
			& & \((t_{13},t_0,t_2,t_2,t_2,t_2,t_3)-
			(t_0,t_1,t_4)-(t_{14},t_{11})-\)\\
			& & \((t_0,t_1,t_4,t_7)-(t_{14},t_0)\)\\
			\hline
			8 & \(t_3-t_{11}-t_{13}-t_{10}-t_5-t_4-t_8-t_{12}\) & \((t_0,t_2,t_2,t_2,t_2,t_3)-t_{11}-(t_0,t_1,t_{13})-\) \\
			& & \((t_0,t_2,t_1,t_4,t_9,t_9,t_9,t_9,t_{10})-\)\\
			& & \((t_0,t_1,t_4,t_9,t_9,t_9,t_9,t_6,t_4,t_5)-t_4-(t_9,t_9,t_9,t_9)-\)\\
			& & \(t_8-(t_0-t_{12})\)\\
			\hline
  		\end{tabular}
	\end{center}
  \label{tab:multtrapresults}
\end{table}

The experimental results in Table \ref{tab:multtrapresults} are given as 8 pairs of test goals and corresponding generated paths. These test goals consist each of 8 implicitly given and dependently defined traps. They are given as a sequence of transitions each of which has an implicit trap defined for it such that the trap constraint consist of a conjunction of trap variables of all the preceding traps in the test goal.
We do not give explicit performance analysis in this section as the average time-wise performance conforms with the results from the previous section.

The test goals 1 - 3 consist of traps of form (i), which means that every trap can be covered immediately after the preceding one. The test goals of form (i) are used to confirm that they are indeed trivial for the combination of \(\chi\)RPT and the RPT. Test goals 4 and 5 both contain one trap of form (ii). It is clearly visible that although these traps can not be immediately covered after preceding ones, \(\chi\)RPT is able to guide the SUT through the necessary intermediate states. The last three test goals 6 - 8 contain traps randomly chosen from all possible traps and therefore contain multiple traps of form (ii). The results again confirm that $\chi$RPT is able to generate near-optimal paths for the test goals of form (ii).

\vspace{-0.6em}
\subsection{Industrial Scale Telecom Billing System}
\label{sec:billingsystem}

In this section we consider an industrial scale telecom billing system. As this example originates from industry, we are unfortunately not permitted to depict it explicitly. Instead, we can only give a description of the given I/O EFSM by its general characteristics which includes 13 locations and 43 transitions between them, 2 input variables having domains [\(0,11\)] and [\(1,32000\)] and 8 state variables having domains [\(0,1\)] (1), [\(0,1000\)] (1) and [\(0,32000\)] (6). On average, transition guards in this model consist of 20 state and input variables connected with logic and arithmetic operations.

The test goal whose analysis is given in Table \ref{tab:billingsystemresults} consists of a sequence of traps corresponding to exceeding the monthly mobile internet usage limit.  As this model is considerably larger both location- and state-wise than the previous models \(M_1\) and \(M_2\), we also use it to compare how different bounds of the RPT off-line algorithm affect \(\chi\)RPT. We consider 4 different search depth bounds - 100, 50, 10 and 2 iterations. The optimal path has length 189 and it is found by the RPT off-line algorithm in 1.3 hours.

\begin{table}
  \caption{Experimental results on the industrial scale telecom billing system}
	\begin{center}
		\begin{tabular}{| l | | c | c | c | c |}
			\hline
			\textbf{Complete off-line RPT} & \multicolumn{4}{|c|}{} \\ 
			Path length (Work time (s)) & \multicolumn{4}{|c|}{189 (4644)} \\ 
			\hline
			\textbf{Bounded off-line RPT} &  & &  & \\
			Depth (Work time (s)) & 100 (2120) & 50 (1086)  & 10 (95) & 2 (16)\\
			\hline
			\textbf{\(\chi\)RPT} & & &  &  \\
			Total path length (Work time (s)) & 230 (6,7) & 255 (17,4) & 275 (17,0) & 1051 (153,4)\\
			\hline
			On-line path length ($\chi$RPT + RPT) & 130 + 100 & 205 + 50 & 265 + 10 & 1049 + 2\\
			\hline
			Avg. time in each state & 0,051 & 0,084 & 0,063 & 0,146 \\
			\hline
			\textbf{Randomized anti-ant} & - & - & - & - \\
			\hline
			\textbf{UPPAAL} & \multicolumn{4}{|c|}{-} \\
			\hline
  		\end{tabular}
	\end{center}
  \label{tab:billingsystemresults}
\end{table}

As one might expect, the anti-ant strategy failed to generate a successful path in reasonable time due to the significantly large search space. Similarly,  UPPAAL also failed to generate a successful trace (and therefore also a test sequence) because the size of the search-space caused the explicit-state model checking to run out of memory. Moreover, UPPAAL's failure was independent of used configuration, e.g., depth-first/breadth-first search, state-space representation and reduction strategies. 

On the other hand, \(\chi\)RPT was able to satisfy the test goal in each of the five cases of different RPT search depth bounds. From Table \ref{tab:billingsystemresults} we can first conclude that the generated path indeed depends on the RPT search depth bounds. This corresponds to the intuition that \(\chi\)RPT is complementing the backward RPT off-line algorithm with a forward on-line algorithm and the farther the RPT generates the reachability constraints, the more information they provide to \(\chi\)RPT. Secondly, we can conclude that the difference between the generated path and the optimal path does not strictly depend on the I/O EFSM size. Although the generated path for the simple counter model \(M_1\) in Sect. \ref{sec:onetrap} was 2 times longer than the optimal, the difference here for bounds 10 to 100 is less than 1.5 times for a significantly larger model. Only for depth 2 the generated path is significantly longer than optimal.

In conclusion, \(\chi\)RPT works efficiently with the given industrial model by generating near-optimal paths time-wise efficiently. As the SUT in this example is a relatively large component of an industrial system, we are able to give empirical backing to the capability of \(\chi\)RPT for handling components of industrial scale systems. Moreover, $\chi$RPT is also able to handle larger models when the RPT off-line algorithm search depth bounds and $\chi$RPT configuration variables are modified accordingly.

\vspace{-0.6em}
\section{Conclusions and Further Work}
\label{sec:discussion}

The motivation behind this research was to improve the scalability and efficiency of on-line test generation from non-deterministic output-observable I/O EFSMs. Although test generation and execution from EFSMs has been studied extensively, on-line test generation from non-deterministic models tends to confine itself to computationally inexpensive but inefficient strategies such as random search or anti-ant. In this paper, we proposed a constraint-based heuristic approach (Heuristic Reactive Planning Tester (\(\chi\)RPT)) for I/O EFSM-based on-line test generation that could be used when the RPT on-line algorithm \cite{vain2011,kaaramees2010} is not applicable. \(\chi\)RPT is based on reachability constraints and properties of the underlying control graph of the I/O EFSM. We compared \(\chi\)RPT with other search strategies such as the RPT \cite{vain2011}, a randomized version of the anti-ant strategy \cite{li2005} and also the ones implemented in the model-checking tool UPPAAL \cite{larsen1997} on a three-variable counter, Inres Initiator and an industrial telecom billing system. 

The models considered in this paper are limited to EFSM models over linear arithmetics. This is an important extension compared to modelling systems using FSMs, but not all SUTs can be easily modelled using only linear arithmetics. All the results are applicable to models over different theories, provided that we have a satisfiability solver, optimization procedure and a function for calculating violations degree of the formulae of the used theory.

We have confined ourselves to only dependency based test goal and trap selection criteria and left additional analysis as further work.  Further research is also needed for the use of \(\chi\)RPT with not connected, hierarchical and distributed I/O EFSMs. Moreover, further work will include improvements to the fitness function computation and stronger trap discarding and ordering conditions.

\vspace{-0.6em}
\section{Acknowledgements}
We thank J\"{u}ri Vain and Kullo Raiend for many valuable comments and discussions.

\vspace{-0.6em}
\bibliographystyle{eptcs}
\bibliography{references,testing}

\end{document}